\documentclass[prb,twocolumn,eqsecnum,showpacs]{revtex4}

\usepackage{graphicx}

\begin{document}  

\title{Pseudo-zero-mode Landau levels and 
collective excitations in bilayer graphene
}
\author{K. Shizuya}
\affiliation{Yukawa Institute for Theoretical Physics\\
Kyoto University,~Kyoto 606-8502,~Japan }

\begin{abstract} 

Bilayer graphene in a magnetic field supports eight zero-energy Landau levels, 
which, as a tunable band gap develops, 
split into two nearly-degenerate quartets separated by the band gap.  
A close look is made into the properties of such an  isolated quartet 
of pseudo-zero-mode levels at half filling 
in the presence of an in-plane electric field and the Coulomb interaction, 
with focus on revealing further controllable features in bilayer graphene.
The half-filled pseudo-zero-mode levels support, via orbital level mixing, 
charge carriers with nonzero electric moment, 
which would lead to field-induced level splitting and 
the current-induced quantum Hall effect.
It is shown that the Coulomb interaction enhances the effect of the in-plane field
and their interplay leads to rich spectra of collective excitations, 
pseudospin waves, accessible by microwave experiments; also 
a duality in the excitation spectra is revealed.

\end{abstract}


\maketitle

\section{Introduction}  

Graphene, a monolayer of graphite, attracts a great deal of attention, 
both experimentally~\cite{NG,ZTSK,ZJS} and
theoretically,~\cite{ZA,GS,PGN,NM,AF}
for its unusual electronic transport, characteristic of 
$\lq\lq$relativistic" charge carriers, massless Dirac fermions. 
Dirac fermions give rise to quantum phenomena reflecting 
the particle-hole picture of the vacuum state, 
such as Klein tunneling,~\cite{KNG} 
and, especially in a magnetic field, 
such peculiar phenomena~\cite{J,NS,Semenoff,Hal,FScs} 
as spectral asymmetry and induced charges,
which are rooted in the chiral anomaly
(i.e., a quantum conflict between charge and chirality conservations). 
Graphene provides a special laboratory to test such consequences of 
quantum electrodynamics.
Actually, the half-integer quantum Hall (QH) effect and 
the presence of the zero-energy Landau level
observed~\cite{NG,ZTSK} in graphene are a manifestation of
spectral asymmetry.

Bilayer graphene is equally
exotic~\cite{NMMKF, MF,KA} as monolayer graphene. 
It has a unique property that the band gap is
controllable~\cite{OBSHR,Mc,CNMPL,MSBM,OHL} 
by the use of external gates or chemical doping;
this makes bilayers richer in electronic properties.
In bilayer graphene interlayer coupling modifies 
the intralayer relativistic spectra 
to yield quasiparticles with a parabolic energy dispersion.~\cite{MF}
The particle-hole structure still remains
in the "chiral" form of a Schroedinger Hamiltonian, 
and there arise eight zero(-energy)-mode Landau levels
(two per valley and spin) in a magnetic field.

Zero-mode Landau levels, specific to graphene, deserve attention 
in their own right. 
They would show quite unusual dielectric response~\cite{dielResp,KSgr}
while they carry normal Hall conductance $e^{2}/h$ per level;
for bilayer graphene direct calculations~\cite{MS}
indicate that the zero-modes show no dielectric response for a zero band gap 
but the response grows linearly with the band gap.
In bilayer graphene, with a tunable band gap,
the zero-mode levels split into two quartets separated by the band gap
at different valleys.
Such an isolated quartet of "pseudo"-zero-mode levels remains
nearly degenerate and, as noted earlier,~\cite{MS} 
the level splitting is enhanced or controlled
by an in-plane electric field or by an injected current;
this opens up the possibility of the current-induced QH effect
for the pseudo-zero-mode sector around half filling,
i.e., at filling factor $\nu=\pm2$.

The purpose of this paper is to
further examine the properties of the pseudo-zero-mode levels, 
especially, coherence and collective excitations 
in the presence of   an external field and the Coulomb interaction.
The pseudo-zero-mode levels at half filling support, 
via orbital level mixing, charge carriers 
with nonzero electric dipole moment,
which is responsible for field-induced level splitting and 
the current-induced QH effect.
Along this line our discussion comes in contact 
with the works of Barlas {\it et al.}~\cite{BCNM}
and Abergel {\it et al.}~\cite{AbF}
who, from the viewpoint of QH ferromagnets, studied 
the interaction-driven QH effect 
in the nearly-degenerate octet of zero-mode levels in bilayer graphene.
Our paper partly extends their analysis by revealing an interesting interplay 
between the in-plane field and Coulomb exchange interaction, 
which leads to rich spectra of collective excitations,
accessible by microwave experiments.
We shall find a duality in the excitation spectra 
and, under certain circumstances, an instability in pseudospin textures.

In Sec.~II we briefly review some basic features of 
the pseudo-zero-mode levels in bilayer graphene.
In Sec.~III we construct a low-energy effective theory 
for the half-filled pseudo-zero-mode sector.
In Sec.~IV we examine the spectrum and collective excitations in it.
In Sec.~V we study the microwave response of collective excitations.
Section~VI is devoted to a summary and discussion.


\section{Bilayer graphene}

Bilayer graphene consists of two coupled hexagonal lattices of 
carbon atoms,
arranged in Bernal $A'B$ stacking.
The electron fields in it are described 
by four-component spinors on the four inequivalent sites 
$(A,B)$ and $(A',B')$ in the bottom and top layers,
and their low-energy features are governed 
by the two inequivalent Fermi points $K$ and $K'$ in the Brillouin zone. 
The intralayer coupling $\gamma_{0} \equiv \gamma_{AB}$ is related to
the Fermi velocity 
$v_{0} = (\sqrt{3}/2)\, a_{\rm L}\gamma_{0}/\hbar \approx 10^{6}$~m/s 
(with $a_{\rm L}=  0.246$nm) in monolayer graphene.
The interlayer couplings $\gamma_{1} \equiv \gamma_{A'B}$ and
$\gamma_{3} \equiv \gamma_{AB'}$ are one-order of magnitude weaker
than $\gamma_{0}$; 
numerically,~\cite{MNEBP} $\gamma_{1} \approx 0.30$~eV, 
$\gamma_{3}\approx 0.10$ eV and  $\gamma_{0} \approx 2.9$~eV.

Actually, interlayer hopping via the $(A',B)$ dimer sites 
modifies the intralayer relativistic spectra
to yield spectra with a quadratic dispersion 
and the characteristic cyclotron energy
$\omega_{c} = 2 v_{0}^{2}/(\gamma_{1}\ell^{2}) 
\approx 3.9\, B[{\rm T}]$~meV, with a magnetic field $B[{\rm T}]$ in Tesla.
The low-energy branches, thereby, are essentially described 
by two-component spinors on the $(A,B')$ sites 
(with the high energy branches being separated 
by a large gap $\sim \gamma_{1}$).
The effective Hamiltonian is written as~\cite{MF}
\begin{eqnarray}
H&=&\!\! \int\! d^{2}{\bf x}\Big[ \psi^{\dag} ({\cal H}_{+}\! - eA_{0}) \psi 
+ \chi^{\dag} ({\cal H}_{-}\! - eA_{0}) \chi \Big], \nonumber\\
{\cal H}_{\xi} &=& {\cal H}_{0}  + {\cal H}_{\rm as}, \nonumber \\
{\cal H}_{0}
&=&  \omega_{\rm c}\left(
\begin{array}{lc}
 & -(a^{\dag})^{2}+\lambda\, a\\
-a^{2} +  \lambda\, a^{\dag} & \\
\end{array}
\right),\nonumber\\
{\cal H}_{\rm as}
&=&\!\!  \xi\, {U\over{2}} 
\left( \! 
\begin{array}{ll}
1 - z\, a^{\dag}a & \\
 &\! - (1 -z\, aa^{\dag}) \! \\ 
\end{array}
\right),
\label{Hbilayer}
\end{eqnarray}
together with coupling to electromagnetic potentials 
$(A_{i}, A_{0})$.
Here, assuming placing graphene in a uniform magnetic field 
$B>0$, we have rescaled the kinetic momenta 
$\Pi_{i} = -i\partial_{i} + e A_{i}$
with the magnetic length $\ell=1/\sqrt{eB}$ and defined
$a= \sqrt{2eB}\, (\Pi_{x} -i \Pi_{y})$ and
$a^{\dag}= \sqrt{2eB}\, (\Pi_{x} +i \Pi_{y})$
so that $[a,a^{\dag}]=1$;
we set 
$A_{i}\rightarrow  B\, (-y,0)$ to supply a strong magnetic field 
$B$ normal to the sample plane.
The field $\psi = (\psi_{A}, \psi_{B'})^{\rm t}$ refers to 
the $K$ valley with ${\cal H}_{+} ={\cal H}_{\xi=1}$ 
while $\chi = (\chi_{B'}, \chi_{A})^{\rm t}$ refers to 
the $K'$ valley with ${\cal H}_{-} ={\cal H}_{\xi=-1}$.
For simplicity we ignore weak Zeeman coupling and 
suppress the electron spin indices.

In ${\cal H}_{0}$ the linear kinetic term, leading to "trigonal warping", represents 
direct interlayer hopping via $\gamma_{3}$, 
with $\lambda = \xi\,  (\gamma_{3}/\gamma_{0}) (\sqrt{2}\, v_{0}/\ell)/\omega_{\rm c} 
\approx \pm 0.3 /\sqrt{B[{\rm T}]}$.

The ${\cal H}_{\rm as}$ 
takes into account a possible layer asymmetry, 
caused by an interlayer voltage $\triangle A_{0}$, 
which leads to a tunable~\cite{OBSHR,Mc,CNMPL} gap 
$U \approx e\triangle A_{0}$ between the conduction and valence bands.
The $O(z a^{\dag}a)$ terms in ${\cal H}_{\rm as}$ represent 
a kinetic asymmetry related to the charge depleted from the dimer sites;
note that it is very weak, with 
$z=2\omega_{\rm c}/\gamma_{1} \approx 0.026\times B[{\rm T}] \ll 1$.

While the linear spectra are lost, the bilayer Hamiltonian 
still possesses a key feature of relativistic field theory, 
the particle-hole (or chiral) structure of the quantum vacuum:
When the tiny $O(z\,  U)$ asymmetry is ignored, 
the spectrum of ${\cal H}_{\xi}$, in general, is symmetric about zero energy 
$\epsilon=0$, apart from possible $\epsilon= \pm U/2$ spectra 
that evolve from the zero-energy modes of ${\cal H}_{0}$.
Indeed, for $\lambda=0$ the spectrum of  ${\cal H}_{\xi}$ 
consists of an infinite tower of Landau levels
$|n, y_{0}\rangle$ of paired positive and negative energies,
\begin{eqnarray}
\epsilon_{n} =  s_{n}\, \omega_{\rm c} \sqrt{|n| (|n| -1) 
+ (U/2\omega_{c})^{2}}  
- {1\over{4}}\, \xi\,z\, U,  
\label{specnonzero}
\end{eqnarray}
labeled by integers $n=\pm 2, \pm3, \dots$, and
$p_{x}$ (or $y_{0} \equiv \ell^{2} p_{x}$); 
$s_{n} \equiv {\rm sgn}\{n\} =\pm 1$ specifies 
the sign of $\epsilon_{n}$.

In addition, there arise nearly-degenerate Landau levels carrying  
the orbital index $|n|=0$ and $|n|=1$, 
with spectrum
\begin{eqnarray}
\epsilon_{|n|=0} &=&  \xi\, U/2,\  
\epsilon_{|n|=1} =  \xi\, (U/2)\, (1-z).    
\label{zeromodespectrum}
\end{eqnarray}
From now on we take, without loss of generality,  
$U> 0$ and denote $n=0_{\pm}$ and $n=\pm 1$ 
to specify these pseudo-zero-mode levels.
There are four such pseudo-zero-mode levels (or two levels per spin) 
at each valley and they reside on different layers;
the $(0_{+},1)$ quartet on $A$ sites at $\xi=1$ valley 
is separated from the the $(0_{-},-1)$ quartet on $B'$ sites
by a band gap $U$.
Each quartet remains degenerate, apart from $O(z\,U)$ fine splitting.

The presence of the pseudo-zero-mode levels
and their double-fold degeneracy (per spin and valley) both have
a topological origin, and are consequences of spectral asymmetry, 
or the nonzero index of the Hamiltonian
${\cal H}_{\xi}|_{U\rightarrow 0}$, 
\begin{eqnarray}
{\rm Index}[{\cal H}_{\xi}|_{U\rightarrow 0}] 
=\int\! d^{2}{\bf x}\, 2\times (eB/2\pi),
\end{eqnarray}
which is tied to the chiral anomaly in 1+1 dimensions.
This degeneracy is unaffected by the presence of 
trigonal warping $\lambda\not= 0$ alone,~\cite{MS}
but is affected by  nontrivial diagonal components in ${\cal H}_{\xi}$.
Indeed, the kinetic asymmetry $\sim z a^{\dag}a$ leads to tiny level splitting
and an electric field $\sim \partial_{i}A_{0}$ can also enhance the splitting.
In other words, the pseudo-zero-mode levels have  
an intrinsic tendency to be degenerate, 
but this at the same time implies that 
their fine structure or the way they get mixed 
depends sensitively on the environment.

The main purpose of the present paper is to study 
such controllable features of the isolated pseudo-zero-mode quartet 
in the presence of external fields and Coulomb interactions. 
We are particularly interested 
in the properties of such a quartet at half filling,
where mixing of the zero-modes leads to nontrivial coherence effects 
and collective excitations.
For definiteness we focus on the $n= (0_{+}, 1)$ sector at $\xi =1$ valley,
i.e., around filling factor $\nu=2$ (or $\nu=1$ when the spin is resolved),
and ignore the presence of 
other levels which are separated by relatively large gaps;   
the $\nu=-2$ case is treated likewise.
We ignore the effect of trigonal warping, which causes
only a negligibly small level splitting~\cite{MS} of $O(z\lambda^{4}) < 1/10^{3}$, 
apart from a common level shift of $O(z\lambda^{2})$.

To project out the $n=(0_{+}, 1)$ sector let us 
make the Landau level structure 
explicit by the expansion
$\psi ({\bf x}, t) = \sum_{n, y_{0}} \langle {\bf x}| n, y_{0}\rangle\, 
\psi_{n}(y_{0},t)$, 
with the field operators obeying 
$\{ \psi_{m}(y_{0},t), \psi^{\dag}_{n}(y_{0},t)\} 
= \delta_{mn} \delta (y_{0}- y'_{0})$.
The charge density 
$\rho_{-{\bf p}}(t) =\int d^{2}{\bf x}\,  e^{i {\bf p\cdot x}}\,\psi^{\dag}\psi$ 
is thereby written as 
\begin{equation}
\rho_{-{\bf p}} = \gamma_{\bf p}\!
\sum_{k, n=-\infty}^{\infty} g_{k n}({\bf p})\int dy_{0}\,
\psi_{k}^{\dag}\, e^{i{\bf p\cdot r}}\,
\psi_{n} , 
\label{chargeoperator}
\end{equation}
where $\gamma_{\bf p} =  e^{- \ell^{2} {\bf p}^{2}/4}$; 
${\bf r} = (r_{x}, r_{y}) = (i\ell^{2}\partial/\partial y_{0}, y_{0})$
stands for the center coordinate with uncertainty 
$[r_{x}, r_{y}] =i\ell^{2}$.
For the $(0_{+},1)$ sector the relevant coefficients are given 
by~\cite{MS} 
 $g_{00}({\bf p})=1$,  $g_{11}({\bf p}) = 1 -  \ell^{2}\, {\bf p}^{2}/2$,
 $g_{10}({\bf p}) = i\, \ell p/\sqrt{2}$, and  $g_{01}({\bf p}) = i\,\ell p^{\dag}/\sqrt{2}$,
with $ p\equiv p_{y}\! +\!ip_{x}$.

Let us put the $\psi_{0_{+}}(y_{0},t)$ and $\psi_{1}(y_{0},t)$ modes
into a two-component spinor $\Psi=(\psi_{0_{+}}, \psi_{1} )^{\rm t}$,
and define the pseudospin operators in the $(0_{+}, 1)$ orbital space,
\begin{eqnarray}
S^{\mu}_{\bf -p} &=&\gamma_{\bf p} \int dy_{0}\, 
\Psi^{\dag}\, {1\over{2}}\,\sigma^{\mu} e^{i{\bf p \cdot r}}\Psi,
\label{Smup}
\end{eqnarray}
where $\sigma^{\mu}= (1, \sigma^{a})$ ($\mu=0 \sim 3$) 
with Pauli matrices $\sigma^{a}$ and $\sigma^{0}=1$.
The charge density $\rho_{-{\bf p}}$ projected to the $(0_{+}, 1)$ sector 
is thereby written as
\begin{equation}
\bar{\rho}_{\bf -p} 
= 2\, ( w^{0}_{\bf p}S^{0}_{\bf -p} + w^{3}_{\bf p} S^{3}_{\bf -p}
 + w^{1}_{\bf p} S^{1}_{\bf -p}+ w^{2}_{\bf p}S^{2}_{\bf -p} ), 
\end{equation}
where $w^{0}_{\bf p} = 1- \ell^{2}{\bf p}^{2}/4$,
 $w^{3}_{\bf p} = \ell^{2}{\bf p}^{2}/4$,
 $w^{1}_{\bf p} = i\ell p_{y}/\sqrt{2}$, and 
$w^{2}({\bf p}) = i\,\ell p_{x}/\sqrt{2}$.
Note that $(S^{2}_{\bf -p},  S^{1}_{\bf -p})^{\rm t}$ acts as a vector 
under rotations about the $z$ axis $\parallel B$.

The Hamiltonian $H$ of Eq.~(\ref{Hbilayer}) is
projected into the $(0_{+}, 1)$ sector to yield
\begin{eqnarray}
\triangle \bar{H}=-\int\! d^{2}{\bf x}\,  e A_{0}\, \bar{\rho} + 
\int\! dy_{0}\, m\, \Psi^{\dag}\sigma^{3}\, \Psi,
\end{eqnarray}
where $m\equiv  z\,  U/4$;
the common energy shift $(1 - z/2) U/2$ 
of the $(0_{+}, 1)$ sector has been isolated from $\triangle \bar{H}$. 
One can equally write $\triangle \bar{H}$ as
\begin{eqnarray}
\triangle \bar{H} 
= 2\sum_{\bf p} \, {\cal P}^{\mu}_{\bf p}\, S^{\mu}_{\bf -p},
\end{eqnarray}
with ${\cal P}^{\mu}_{\bf p}$ being the Fourier transform 
of ${\cal P}^{\mu}({\bf x},t)$,
\begin{eqnarray}
{\cal P}^{0}&=& - m -eA_{0} 
- \textstyle{1\over{4}}\, e\ell^{2} \nabla\!\cdot\!{\bf E}, \nonumber\\
{\cal P}^{3}&=& m - \textstyle{1\over{4}}\,  
e\ell^{2} \nabla\!\cdot\! {\bf E} ,  \nonumber\\
({\cal P}^{2},  {\cal P}^{1}) &=& {e\ell\over{\sqrt2}}\,( E_{x}, E_{y}),  
\end{eqnarray}
where ${\bf E}_{\parallel} =(E_{x}, E_{y}) = -\partial_{\bf x}A_{0}$ 
denotes the in-plane electric field.

The Coulomb interaction also simplifies via projection. 
The pseudo-zero-modes at each valley essentially   
lie on the same layer, apart from a negligibly small admixture 
of $O(\omega_{c}/\gamma_{1}) \sim 10^{-3}$. 
One may  thus retain only the intralayer interaction for $\Psi$,
\begin{equation}
\bar{H}^{\rm C} 
= {1\over{2}} \sum_{\bf p}
v_{\bf p} :\bar{\rho}_{\bf -p}\, \bar{\rho}_{\bf p}: ,
\label{Hcoul}
\end{equation}
where 
$v_{\bf p}= 2\pi \alpha/(\epsilon_{\rm b} |{\bf p}|)$ is 
the Coulomb potential with the fine-structure constant 
$\alpha = e^{2}/(4 \pi \epsilon_{0}) \approx 1/137$ and 
the substrate dielectric constant $\epsilon_{\rm b}$;
$\sum_{\bf p} =\int d^{2}{\bf p}/(2\pi)^{2}$. 
The normal-ordered charges in $\bar{H}^{\rm C}$ are rewritten as 
\begin{eqnarray}
:\bar{\rho}_{\bf -p}\, \bar{\rho}_{\bf p}:
&=& \bar{\rho}_{\bf -p}\, \bar{\rho}_{\bf p} - \triangle, \nonumber\\
\triangle &=& 2\, \gamma_{\bf p}^{2}\,  
\Big\{ |w^{\mu}_{\bf p}|^{2}\, S^{0}_{\bf 0} 
+ 2 w^{0}_{\bf p}w^{3}_{\bf p}\, S^{3}_{\bf 0} \Big\}\ \ \ \ 
\label{triangle}
\end{eqnarray}
under a symmetric integration over ${\bf p}$.

The pseudospin operators $S^{\mu}_{\bf p}$ obey 
the SU(2)$\times W_{\infty}$ algebra,~\cite{GMP,MoonMYGM}
\begin{eqnarray}
&&[S^{a}_{\bf p},S^{b}_{\bf k}]= 
 c(p,k)\,i\epsilon^{abc}\, S^{c}_{\bf p+k}
- i\, s(p,k)\, \delta^{ab}\,S^{0}_{\bf p+k}, \nonumber\\
&&[S^{0}_{\bf p}, S^{0}_{\bf k}]
= -i\, s(p,k) \, S^{0}_{\bf p+k} , \nonumber\\
&&[S^{0}_{\bf p}, S^{a}_{\bf k}]
=[S^{a}_{\bf p}, S^{0}_{\bf k}]
= -i\, s(p,k) \, S^{a}_{\bf p+k}, 
\label{chargealg}
\end{eqnarray}
where $(a,b)$ runs over $1 \sim 3$, and 
\begin{eqnarray}
s(p,k) = \sin \Big({\ell^{2}{\bf p}\!\times\! {\bf k}\over{2}}
\Big)\ e^{ \ell^{2} {\bf p\cdot k}/2}; 
\end{eqnarray}
${\bf p} \times {\bf k} \equiv \epsilon^{ij}p_{i}k_{j}
= p_{x}k_{y}- p_{y}k_{x}$; for $c(p,k)$ set $\sin (\cdots) 
\rightarrow \cos (\cdots)$ in $s(p,k)$.

\section{Pseudospin textures}

In this section we study the properties of the pseudo-zero-mode levels
at half filling, using the projected Hamiltonian  
$\bar{H} \equiv \triangle \bar{H} + \bar{H}^{\rm C}$
and the charge algebra~(\ref{chargealg}), 
with focus on orbital mixing of the zero-modes. 
Let us suppose that such a half-filled state 
is given by a classical configuration where the pseudospin points in  
a fixed direction in pseudospin space, i.e., 
$S^{a}_{\bf p=0} = {1\over{2}}\,
N_{e}\, n^{a}$ and $n^{a}n^{a} =1$ with the total number 
of electrons $N_{e}= 2\, S^{0}_{\bf p=0}$.

Note that $n^{3}=1$ corresponds to the filled $n=0_{+}$ level 
with the vacant $n=1$ level while 
$n^{3} = -1$ represents the filled $n=1$ level.
The direction ${\bf n}=(n^{1},n^{2},n^{3})$ would, in general, 
vary in response to the external field $A_{0}$, and,
as ${\bf n}$ tilts from $n^{3}=\pm 1$, 
the $n=0_{+}$ and $n=1$ levels start to mix.
For selfconsistency we assume that  $A_{0}$ represents a uniform in-plane electric 
field ${\bf E}_{\parallel} = -\partial_{\bf x}A_{0}$ and that it leads to 
a homogeneous state $|G\rangle$ of uniform density $\rho_{0}= \nu/(2\pi \ell^{2})$
(with filling factor $\nu=2$ for the spin-degenerate $\nu=2$ state
or $\nu=1$ for the spin-resolved $\nu=1$ state).
We thus consider all classical configurations 
with $\langle G|S^{a}_{\bf p=0}|G \rangle = {1\over{2}}\,
N_{e}\, n^{a}$ and single out the ground state or the associated  $n^{a}$ 
by minimizing the energy $\langle G| \bar{H} |G \rangle$.
For $n^{a}$ we use the parametrization 
$n^{1}= \sin\theta\, \cos \phi$, $n^{2}= \sin\theta\, \sin \phi$, 
$n^{3}= \cos \theta$, with $-\pi < \theta \le \pi$ and $0\le \phi \le  \pi$.
We denote expectation values 
$\langle G| {\cal O} |G \rangle \equiv \langle {\cal O} \rangle$
for short.

Let us first substitute the charge and pseudospin,
$\langle S_{\bf p}^{0}\rangle = {1\over{2}} \rho_{0}\, \delta_{\bf p,0}$ 
and
$\langle S_{\bf p}^{a}\rangle 
= {1\over{2}} \rho_{0}\,  n^{a}\,  \delta_{\bf p,0}$, 
into $\triangle \bar{H}$.
(Here $\delta_{\bf p,0} =  (2\pi)^{2}\, \delta^{2} ({\bf p})$,
and  $\delta_{\bf 0,0} =\int d^{2}{\bf x}$ equals the total area so that
$N_{e}=\rho_{0}\,\delta_{\bf 0,0}$.)
This yields the response of the classical state 
$|G\rangle$ to an external probe,
$\langle \triangle \bar{H}\rangle 
= \rho_{0}\int d^{2}{\bf x}\,  n^{\mu} {\cal P}^{\mu}$ with
\begin{eqnarray}
n^{\mu} {\cal P}^{\mu}&=& - e A_{0} + m\,  (\cos \theta -1)
- {\bf E}_{\parallel}\cdot {\bf d}_{e}, \nonumber\\ 
{\bf d}_{e}&=& -{e\ell\over{\sqrt{2}}}\, (n^{2}, n^{1}) 
= -{e\ell\over{\sqrt{2}}}\, \sin\theta\,  \hat{\bf n}_{\parallel},
\end{eqnarray}
where 
$\hat{\bf n}_{\parallel}=(\hat{n}_{x}, \hat{n}_{y}) 
= (\sin \phi, \cos \phi)$
is an in-plane unit vector.

 Note that the half-filled pseudo-zero-mode state 
has an in-plane electric dipole moment ${\bf d}_{e}$
of strength $(e\ell/\sqrt{2})\, |\sin\theta|$ per electron, 
proportional to the in-plane component $(n^{2}, n^{1})$ of the pseudospin.
Mixing of the $n=0_{+}$ and $n=1$ modes gives rise to this dipole 
and its in-plane direction $\hat{\bf n}_{\parallel}$ is related to 
the relative phase between them;
i.e., under U(1) transformations 
$\Psi=(\psi_{0_{+}}, \psi_{1} )^{\rm t} \rightarrow
e^{-i\phi_{\rm rel}\, \sigma^{3}/2}\,\Psi$,
$\hat{\bf n}_{\parallel}$ rotates, 
\begin{equation}
(S^{1}+iS^{2}) \rightarrow e^{i\phi_{\rm rel}}\, (S^{1}+iS^{2})\ 
{\rm or\ }
\phi \rightarrow \phi + \phi_{\rm rel}.
\label{rotation}
\end{equation}
This tells us that a change in relative phase caused by a change 
in direction of the dipole is physically observable.

In the absence of the Coulomb interaction, $n^{a}$ naturally 
points in the $-{\cal P}^{a}$ direction and, as a result, 
an in-plane field, coupled to the the electric dipole, 
works to enhance the pseudo-zero-mode  level splitting,~\cite{MS}
\begin{equation}
\triangle \epsilon  = 2\sqrt{m^{2} 
+ e^{2}\ell^{2} {\bf E}_{\parallel}^{2}/2}.
\end{equation}
We shall discuss below how the Coulomb interaction modifies this.

The calculation of the Coulomb energy 
$\langle G| \bar{H}^{\rm C} |G\rangle \equiv \langle \bar{H}^{\rm C} \rangle$ 
requires the knowledge of pseudospin structure factors 
$\langle G|S^{\mu}_{\bf p} S^{\nu}_{\bf q}|G \rangle \equiv  
\langle S^{\mu}_{\bf p} S^{\nu}_{\bf q} \rangle$, which, 
for the present half-filled state with pseudospin $\propto n^{a}$, are given by
\begin{equation}
\langle S^{\mu}_{\bf p} S^{\nu}_{\bf q} \rangle 
= {1\over{4}}\, \rho_{0}
\Big[ s^{\mu\nu}\,  \gamma_{\bf p}^{2} \delta_{{\bf p+ q,0}} 
+ n^{\mu}n^{\nu}\rho_{0} \delta_{{\bf p,0}}\,  \delta_{{\bf q,0} }
\Big],
\label{SpSq}
\end{equation}
where $\mu$ and $\nu$ run from 0 to 3, with $n^{0}= 1$ 
and
$\{n^{a}\} = (n^{1}, n^{2}, n^{3})$;
$(s^{\mu\nu})\dag = s^{\nu\mu}$ with $s^{\mu0} =0$ and 
\begin{eqnarray}
s^{33}&=& \sin^{2}\theta,\nonumber\\ 
s^{11}&=& \cos^{2}\theta
\cos^{2}\phi\ + \sin^{2} \phi,\nonumber\\ 
s^{22}&=& \cos^{2}\theta
\sin^{2}\phi\ + \cos^{2} \phi,\nonumber\\ 
s^{31}&=& -\sin \theta \cos\theta\cos \phi + i \sin \theta \sin\phi, \nonumber\\ 
s^{32}&=& -\sin \theta \cos\theta \sin \phi - i \sin \theta \cos\phi, \nonumber\\ 
s^{12}&=& -\sin^{2} \theta \sin \phi \cos \phi + i \cos \theta \cos 2\phi.
\end{eqnarray}
See Appendix A for details.
Actually, the normal-ordered correlation functions take simpler form
\begin{equation}
\langle\ :\! S^{\mu}_{\bf p} S^{\nu}_{\bf q}: \rangle 
= -{1\over{4}}\, \rho_{0}\, n^{\mu}n^{\nu}\, 
(\gamma_{\bf p}^{2} \delta_{{\bf p+ q,0}} 
-\rho_{0}\, \delta_{{\bf p,0}}\,  \delta_{\bf q,0} ),
\label{normSaSb}
\end{equation}
with which one can cast the Coulomb energy in the form 
\begin{eqnarray}
\langle \bar{H}^{\rm C} \rangle = -{1\over{2}}\,N_{e}\sum_{\bf p}v_{\bf p}\,
e^{-q^{2}/2}\, C_{\theta}(q^{2}/4), 
\label{CoulombEnergy}
\end{eqnarray}
where $q^{2}\equiv \ell^{2}\, {\bf p}^{2}$ and $C_{\theta}(x) 
= 1 - x(1-x)\, (1- \cos \theta)^{2}$.
From $\langle \bar{H}^{\rm C} \rangle$ we have omitted a constant
$(N_{e}/2)\,\rho_{0}\,  v_{\bf p\rightarrow 0}$ which is removed when 
the neutralizing positive background is taken into account.
Integrating over ${\bf p}$ yields a typical scale of
the Coulomb exchange energy, 
\begin{equation}
V_{1} = \sum_{\bf p}v_{\bf p}\, e^{-\ell^{2}{\bf p}^{2}/2} 
= {\alpha\over{\epsilon_{b}\, \ell}}\, \sqrt{{\pi \over{2}}},
\end{equation}

The effective Hamiltonian $H_{\rm eff}= \langle \triangle\bar{H} 
+ \bar{H}^{\rm C}\rangle$ is conveniently written in ${\bf x}$ space as
$H_{\rm eff}=  \rho_{0}\int d^{2}{\bf x}\, {\cal H}_{\rm eff}$ with
\begin{eqnarray}
 {\cal H}_{\rm eff}
&=& - e A_{0} + E(\theta),   \nonumber\\ 
E(\theta)
&=& - {1\over{2}}\, V_{1}  + m\, (\cos \theta -1) + {\cal E}\,  \sin\theta 
\nonumber\\ 
&&
+ {1\over{32}}\,V_{1}\, (1- \cos \theta)^{2},
\label{Etheta}
\end{eqnarray}
where ${\cal E} = (e\ell/\sqrt{2})\, 
{\bf E}_{\parallel}\cdot \hat{\bf n}_{\parallel}$.
In the present notation 
(with $\rho_{0}\int d^{2}{\bf x} \rightarrow  N_{e}$)
${\cal H}_{\rm eff}$ stands for energy per electron 
in state $|G\rangle$ with pseudospin $\propto {\bf n}$.
The Coulomb correlation energy $\propto V_{1}$ consists of 
a negative uniform component $-V_{1}/2$ [relative to the zero of energy
$(U/2)\, (1- z/2)$] and a polarization dependent 
component which alone favors $\theta=0$, 
i.e., the filled $n=0_{+}$ level, and which varies continuously 
by an amount $\triangle E_{c}= (1/8)\, V_{1}$ as ${\bf n}$
sweeps in pseudospin space.
The Coulomb interaction thus significantly enhances 
the pseudo-zero-mode level splitting.

The stable configuration of the half-filled zero-mode state $|G\rangle$
is determined by minimizing $E(\theta)$
with respect to ${\bf n}$, or $(\theta, \phi)$.
Obviously $E(\theta)$ depends on $\phi$ through ${\cal E}$,
which attains a maximum when
${\bf n}_{\parallel} \parallel {\bf E}_{\parallel}$,
or ${\cal E} =  e\ell |{\bf E}_{\parallel}|/\sqrt{2}$.
Accordingly it is    convenient, without loss of generality, to suppose that 
the in-plane field ${\bf E}_{\parallel}$ and ${\bf n}_{\parallel}$ 
lie along the $y$ axis,
${\cal E} = e \ell E_{y}/\sqrt{2} \ge 0$ and $\phi=0$.
With this choice the "1", "2" and "3" axes in pseudospin space coincide 
with the $y$, $x$ and $-z$ axes in real space, respectively.
We adopt this choice and set 
$(n^{1}, n^{2}, n^{3}) = ( \sin \theta, 0,\cos \theta)$ in what follows.

One can now look for possible ground-state configurations 
by writing down a phase diagram as a function 
of $m$ and ${\bf E}_{\parallel}$.
For clarity of exposition we leave it for a later stage and 
here study collective excitations over a given ground state 
$|G\rangle|_{\bf n}$.
We focus on a special class of low-energy collective excitations,
pseudospin waves, 
that are rotations about the energy minimum $|G\rangle|_{\bf n}$ 
in pseudospin space.

As is familiar from the case of quantum Hall ferromagnets,~\cite{MoonMYGM}
such a collective state is represented as a texture state
\begin{eqnarray}
|\tilde{G}\rangle = e^{-i{\cal O}} |G\rangle|_{\bf n},
\end{eqnarray}
where the operator $e^{-i{\cal O}}$ with
\begin{eqnarray}
{\cal O}= \sum_{\bf p}\gamma_{\bf p}^{-1}\,\Omega^{a}_{\bf p}\, S^{a}_{\bf -p}
\end{eqnarray}
locally tilts the pseudospin from ${\bf n}$ to $\tilde{\bf n}_{\bf p}$ 
by small angle $\Omega_{\bf p} \sim {\bf n} \times \tilde{\bf n}_{\bf p}$.
Repeated use of the charge algebra~(\ref{chargealg})
then allows one to express the texture-state energy 
$\langle \tilde{G}|\bar{H}|\tilde{G}\rangle
=\langle G|e^{i{\cal O}}\bar{H} e^{-i{\cal O}}|G\rangle$
in terms of the structure factors in Eq.~(\ref{SpSq}), 
and this yields an effective Hamiltonian
as a functional of $\Omega^{a}_{\bf p}$ or its ${\bf x}$-space 
representative $\Omega^{a} ({\bf x},t)$.

The angle variables $\Omega^{a}({\bf x},t)$ may be normalized 
so that $\sum_{a}(\Omega^{a})^{2}=1$ classically. 
As we shall see below, the effective theory is expressed in terms of
the following components of $\Omega^{a}$,
\begin{eqnarray}
\eta &\equiv& \Omega^{1}\cos \theta 
- \Omega^{3}\sin \theta, \nonumber\\ 
\zeta &\equiv& \Omega^{2},
\end{eqnarray}
along the two orthogonal axes
(i.e., the tilted "1" axis and the "2" axis)
perpendicular to the pseudospin ${\bf n}$,
with normalization 
$(n^{a}\Omega^{a})^{2} +\eta^{2} + \zeta^{2}=1$. 
(Here we have chosen $\phi=0$ and set 
$\{n^{a}\}= (\sin \theta, 0,\cos \theta)$,
as remarked above; the $\phi \not =0$ case, if needed, 
is readily recovered.~\cite{fnOmega})
Actually they refer to the following induced pseudospin components
in the excited state $|\tilde{G}\rangle$,
\begin{eqnarray}
\eta &\sim&  - \gamma_{\bf p}^{-1}\, 
\langle \tilde{G}|\, S^{2} \, |\tilde{G}\rangle,
\nonumber\\ 
\zeta&\sim& \gamma_{\bf p}^{-1}\, \langle \tilde{G}|\,
S^{1} \cos \theta - S^{3} \sin \theta \, |\tilde{G}\rangle, 
\end{eqnarray}
as seen from the induced pseudospin 
$\delta \langle \tilde{G}|S^{a}_{\bf p} |\tilde{G} \rangle 
\approx (\rho_{0}/2) \gamma_{\bf p}
\epsilon^{abc}\Omega^{b}_{\bf p}\, n^{c}$.

We expand $\langle \tilde{G}| \bar{H} |\tilde{G}\rangle$ 
to second order in $\Omega$ and retain 
all powers of derivatives acting on $\Omega$ to study the spectrum 
over a wide range of wavelengths. 
The calculation is outlined in Appendix B. The result is
\begin{eqnarray}
\langle \tilde{G}| \bar{H} |\tilde{G}\rangle \!\!
&=&\! \rho_{0}\int d^{2}{\bf x}\, ({\cal H}_{\rm eff} + {\cal H}_{\rm coll}),
\nonumber\\ 
{\cal H}_{\rm coll} 
&=&  
{1\over{2}}\, \eta\, \Gamma_{\eta}\, \eta
+ {1\over{2}}\,  \zeta\, \Gamma_{\zeta}\, \zeta 
 + \zeta\, W_{\bf p}\, \eta + {\cal H}_{\rm ch}+ \delta {\cal H}, \nonumber\\ 
\Gamma_{\eta} &=& - {\cal E}/\sin\theta + F_{\bf p},\nonumber\\ 
\Gamma_{\zeta} &=& E''(\theta) + G_{\bf p}, \nonumber\\ 
\delta {\cal H} &=& E'(\theta)\, 
\{  \zeta +   \Omega^{1} \eta/ (2\sin\theta) \};
 \label{Hcoll}
\end{eqnarray}
$E'(\theta)= dE(\theta)/d\theta$, etc.
Here
\begin{eqnarray}
F_{\bf p} &=& {V_{c}\over{2}}\, \Big[
{1\over{2}}\sqrt{\pi\over{2}}
+ \xi_{q} 
+P_{-}\,
\lambda_{q} \Big],
\nonumber\\ 
G_{\bf p}&=& {V_{c}\over{2}}\, \Big[{1\over{2}} \sqrt{\pi\over{2}}
+ \xi_{q}
-  b_{q}\, \sin^{2}\theta- P_{-}\, \lambda_{q}\,
\cos^{2}\theta \Big],
\nonumber\\ 
W_{\bf p}&=&\!\! -{V_{c}\over{2}}\, 
\Big[ 2\, {p_{x} p_{y}\over{{\bf p}^{2}}}\,
\lambda_{q}\, \cos\theta
+i {\ell  p_{x}\over{\sqrt{2}}}\, \tau_{q}\,  \sin\theta \Big],
\ \ \ \ \ \
\label{FpGpWp}
\end{eqnarray}
with $V_{c}= \alpha/(\epsilon_{b}\, \ell)$, 
$P_{-}=(p_{x}^{2}- p_{y}^{2})/{\bf p}^{2}$ and 
\begin{eqnarray}
\xi_{q}&=&  \nu\,e^{-q^{2}/2}\, 
{q\over{2}}
- \int_{0}^{\infty} \!\! d z\, 
e^{-z^{2}/2}\, \Big( 1- {1\over{2}}\, z^{2} \Big)\, J_{0}(z q), \nonumber\\ 
\lambda_{q}&=&  \nu\,e^{- q^{2}/2}\,  {q\over{2}}
-  {1\over{2}} \int_{0}^{\infty}\!\! d z\, 
e^{- z^{2}/2}\,z^{2}\,  J_{2}(z q),
\nonumber\\ 
b_{q}&=& 
\nu\,e^{- q^{2}/2}\,
( q/2 -   q^{3}/8) \nonumber\\ 
&&
+  {1\over{8}}\int_{0}^{\infty}\!\! d z\, 
e^{- z^{2}/2}\, (1 - 4z^{2} + z^{4})\, 
J_{0}(z q), \nonumber\\ 
\tau_{q} &=& \nu\, e^{- q^{2}/2}\, q/2 \nonumber\\
&&+{2\over{q}} \int_{0}^{\infty}\!\! d z\, 
e^{- z^{2}/2}\, z\, \Big(1- {1\over{4}}\, z^{2} \Big)\, J'_{0}(z q),
\label{integrals}
\end{eqnarray}
where $q=  \ell |{\bf p}|$;
substitution ${\bf p} \rightarrow  -i \nabla$ is understood 
in the ${\bf x}$ representation.
For ${\bf p} \rightarrow 0$,
$\xi_{q} \rightarrow - (1/2)\, \sqrt{\pi/2}$,
$(\lambda_{q}, b_{q} ) \rightarrow 0$
and $\tau_{q} \rightarrow - (1/4)\, \sqrt{\pi/2}$
while they all tend to zero for   ${\bf p} \rightarrow \infty$.
See Appendix C for more explicit forms of the integrals 
involved in these functions.

In ${\cal H}_{\rm coll}$,  ${\cal H}_{\rm ch}$ refers 
to a topological charge 
(to be detected with a constant potential $A_{0}$) 
\begin{equation}
\rho_{0}\! \int\!\! d^{2}{\bf x}\, {\cal H}_{\rm ch}\! 
= -e \int\!\! d^{2}{\bf x}\, A_{0}\, {\nu\over{8\pi}}\, 
 \epsilon^{abc}\epsilon^{ij}\, (\partial_{i}\Omega^{a})
 (\partial_{j}\Omega^{b})\, n^{c},
\end{equation}
where $\epsilon^{xy}=1$ and $\epsilon^{123}=1$.
With $n^{c}$ promoted to $\Omega^{c}$,
$ \int d^{2}{\bf x}\, {\cal H}_{\rm ch}$ involves the winding number, 
which implies~\cite{MoonMYGM} that 
possible topologically-nontrivial semiclassical excitations (Skyrmions) 
associated with $\Omega^{a}$,
in general, carry electric charge of integral multiples of $\nu\, e$. 
Note also that $\delta {\cal H}$, involving a term linear 
in $\zeta$, disappears for a stable-state configuration 
for which $E'(\theta)=0$.

The the kinetic term for $\Omega^{a}$ is supplied 
from the electron kinetic term as Berry's phase, 
\begin{eqnarray}
\langle \tilde{G}| i\partial_{t}| \tilde{G}\rangle
&=&  - {1\over{4}}\,\rho_{0}\,\sum_{\bf k} n^{c}\epsilon^{abc}
\Omega_{\bf -k}^{a} 
\partial_{t}\Omega_{\bf k}^{b} \nonumber\\ 
&=& {\rho_{0}\over{2}}  \int d^{2}{\bf x}\, 
\zeta\, \partial_{t}\eta,
\end{eqnarray}
to $O(\Omega \Omega)$, apart from surface terms. 
This shows that $\zeta$ is canonically conjugate
to $(\rho_{0}/2)\,\eta$. 
One can now write the effective Lagrangian 
for the collective excitations as
\begin{eqnarray}
L &=& {1\over{2}}\, \zeta\, \partial_{t}\eta 
-{\cal H}_{\rm coll}[\eta, \zeta].
\label{Lcoll}
\end{eqnarray}
Note here that this Lagrangian is written as
\begin{equation}
\rho_{0}\int d^{2}{\bf x}\, L=\langle G|e^{i{\cal O}}\, 
(i \partial_{t} -\bar{H}) e^{-i{\cal O}}|G\rangle.
\end{equation}
This representation realizes and systematizes 
the single-mode approximation~\cite{GMP}  
(SMA) within a variational framework.~\cite{KSsma}
The present theory thus embodies nonperturbative aspects of the SMA.

Upon elimination of $\zeta$, Eq.~(\ref{Lcoll}) leads to an alternative 
form of the effective Lagrangian for $\eta$ as follows:
\begin{eqnarray}
L_{\Phi} &=& {1\over{2}}\, (\partial_{t} \Phi)^{2} 
- {1\over{2}}\, \Phi\,  (M_{\bf p})^{2}\, \Phi,
\nonumber\\ 
M_{\bf p}
&=& 2\, \sqrt{\Gamma_{\eta}\, \Gamma_{\zeta} - |W_{\bf p}|^{2}},
\label{LMp}
\end{eqnarray}
where we have set $\Phi = (1/2) \, (\Gamma_{\zeta})^{-1/2}\, 
\eta$.

 The spectrum $M_{\bf p}$ of collective excitations is 
in general anisotropic in ${\bf p}$ at low energies and depends 
critically on the stable-state configuration ${\bf n}$.
In particular, the leading long-wavelength correction in 
$ {\cal H}_{\rm coll}$ starts with 
the direct Coulomb interaction of $O({\bf p})$,
which leads to the spectrum,
\begin{eqnarray}
M_{\bf p}
&\approx& \sqrt{ (2\, \kappa_{\eta}\, \kappa_{\zeta})^{2} + V_{c}\,  
{\nu\, \ell\over{|{\bf p}|}}\,\{ {\kappa}_{\eta}^{2}\, \cos^{2}\theta\,p_{y}^{2} 
+ \kappa_{\zeta}^{2}\, p_{x}^{2}  \} },\nonumber\\ 
\kappa_{\eta}^{2} &\equiv& \Gamma_{\eta}|_{{\bf p}=0}
= - {\cal E}/\sin\theta, \nonumber\\ 
\kappa_{\zeta}^{2} &\equiv& \Gamma_{\zeta}|_{{\bf p}=0} = E''(\theta).
\label{LWspec}
\end{eqnarray}
The  excitation gap at zero wave vector is thus given by
$M_{{\bf p}\rightarrow 0}= 2\kappa_{\eta}\kappa_{\zeta}$.
In contrast, $M_{\bf p}$ recovers isotropy and
the standard excitonic behavior~\cite{KH}
at short wavelengths,
\begin{equation}
M_{{\bf p}\rightarrow \infty} 
\approx V_{1}/2 + \kappa_{\eta}^{2}+ \kappa_{\zeta}^{2},
\label{Minf}
\end{equation}
with $\Gamma_{\eta} \rightarrow V_{1}/4 + \kappa_{\eta}^{2}$,
$\Gamma_{\zeta} \rightarrow V_{1}/4 + \kappa_{\zeta}^{2}$ and
$W_{\bf p} \rightarrow0$ for ${\bf p}\rightarrow \infty$;
$V_{1}= V_{c}\sqrt{\pi/2}$.

It is worth noting here that the Coulomb interaction alone 
yields $\kappa_{\eta}^{2}=0$, i.e., a flat direction in $(\eta,\zeta)$ space.
This implies that, unlike in ordinary bilayer QH systems, there is 
no cost of interlayer capacitance energy for the pseudo-zero-mode sector
which essentially resides in the same layer. 
Coherence is thus easier to form in 
this sector of bilayer graphene.

If we set $\theta\rightarrow 0$, our $M_{\bf p}$ precisely reproduces
the excitation spectrum derived in Ref.~[\onlinecite{BCNM}] 
by assuming spatial isotropy; 
actually, the effective Hamiltonian of Ref.~[\onlinecite{BCNM}] is apparently 
different from our ${\cal H}_{\rm coll}$  
but the spectrum turns out to be the same.
It is our use of general textures ${\bf n}$ 
that allows ${\cal H}_{\rm coll}$ 
to handle spatially anisotropic situations as well.

\section{possible ground states and collective excitations over them}

In this section we study the spectrum of 
the half-filled pseudo-zero-mode state  
and the associated collective excitations.
Let us first gain a rough idea of the strengths of 
$m$ and ${\bf E}_{\parallel}$ relative to 
the Coulomb correlation energy $\triangle E_{c}$.
A naive estimate 
\begin{eqnarray}
\triangle E_{c} = {1\over{8}}\, \sqrt{{\pi\over{2}}}\, 
{\alpha\over{\epsilon_{b}\,\ell}}\approx 2.2\, \sqrt{B[{\rm T}]}\
{\rm meV}
\label{deltaEc}
\end{eqnarray}
with a typical value $\epsilon_{b} \sim 4$ indicates
that the bare Coulomb interaction is apparently sizable,
as compared with the basic Landau gap  
$\omega_{\rm c} \approx 3.9\,  B[{\rm T}]$ meV.
We remark that  Eq.~(\ref{deltaEc}) is likely to 
overestimate  $\triangle E_{c}$.
Actually $\triangle E_{c}$ is written as an integral of the form
\begin{eqnarray}
\triangle E_{c} \sim 2\sum_{\bf p}v_{\bf p}\,
e^{-q^{2}/2} (q^{2}/4) (1-q^{2}/4)
\end{eqnarray}
with $q= \ell |{\bf p}|$, and 
the main contribution comes from the momentum region 
$|{\bf p}|\ell \sim 1$, where 
the Coulomb interaction is very efficiently 
weakened [$v_{\bf p} \rightarrow v_{\bf p}/\epsilon({\bf p})$], 
as indicated by a random-phase-approximation study~\cite{MS}
of the dielectric function $\epsilon ({\bf p})$.
This screening effect essentially comes
from vacuum (Dirac-sea) polarization, specific to graphene.
It may effectively be taken care of 
by setting $\epsilon_{b}\rightarrow \epsilon_{b}\epsilon_{\rm sc}$;
a simple estimate~\cite{fnvac} 
gives $\epsilon_{\rm sc} \sim 9$ at $B\sim$1T
and $\epsilon_{\rm sc} \sim 3.6$ at $B\sim$10T.

The ratio of the intrinsic zero-mode level gap 
$2m =0.013\,  B[{\rm T}]\, U$ to $\triangle E_{c}$ 
is generally small,
\begin{equation}
m/\triangle E_{c} \approx 3\, \epsilon_{\rm sc}\times 10^{-3}\,  \sqrt{B[{\rm T}]}\, 
U[{\rm meV}],
\end{equation}
for a band gap $U$ of $O(1\, {\rm meV})$, and increases with $B$ and $U$.
For the in-plane field ${\cal E}= e \ell |{\bf E}_{\parallel}|/\sqrt{2}$
the ratio
\begin{equation}
{\cal E}/\triangle E_{c} \approx 0.9\, \epsilon_{\rm sc}
\times 10^{-3}\,   E[{\rm V/cm}]/B[{\rm T}]
\end{equation}
is on the order of 10\% for $E =|{\bf E}_{\parallel}|=$ 10 V/cm at $B= 1$T
with $\epsilon_{\rm sc} \sim 9$.
Note that ${\cal E} \approx m$ for  
${\cal E} \approx 3.6\ {\rm  V/cm} \times U[{\rm meV}]\, B[{\rm T}]^{3/2}$ .

To determine the orbital configuration of 
the half-filled zero-mode state $|G\rangle$
one has to look for the minimum of $E(\theta)$, 
$E_{\rm min} = E(\theta_{\rm min})$ with $E'(\theta_{\rm min})=0$.

Let us begin with the case where ${\bf E}_{\parallel}$ is absent.
It is clear from $E(\theta)$ of Eq.~(\ref{Etheta}) 
that the Coulomb correlation favors $\theta=0$ 
while the intrinsic asymmetry $m>0$ alone favors 
$\theta=\pi$;
$\triangle E_{c}$ and $m$ thus compete.

(i) In case $m > \triangle E_{c}$ (although rather unrealistic), 
one finds $E_{\rm min} = - ( 2m -\triangle E_{c})$ 
and $\kappa_{\eta}^{2} =\kappa_{\zeta}^{2} = m- \triangle E_{c}$
at $\theta = \pi$.
The collective excitations have a finite energy gap
$M_{{\bf p}=0} = 2\, (m- \triangle E_{c}) > 0$
and the spectrum is isotropic, 
reflecting the rotational invariance of the bilayer system 
about the $z$ axis $\parallel B$, or more explicitly, 
invariance of $\triangle \bar{H} + \bar{H}^{\rm C}$ (for $A_{0}=0$)
under $U(1)$ rotations in Eq.~(\ref{rotation}).

(ii) On the other hand, for $0 < m < \triangle E_{c}$, one finds
$E_{\rm min} = -  \triangle E_{c}\, (m/\triangle E_{c})^{2}$ 
at $\theta =\pm\theta_{\rm min}$ with 
$\sin^{2} (\theta_{\rm min}/2) = m/\triangle E_{c}$;
$\theta_{\rm min}$ varies from 0 to $\pi$ with increasing $m$. 
Here we encounter a somewhat strange situation.
Note that, for $\theta \not = 0,\pm \pi$, the rotational 
invariance is spontaneously broken. 
Indeed, one finds $\kappa_{\eta}^{2} =0$ and
$\kappa^{2}_{\zeta} =  2 m\, (1-  m/\triangle E_{c})$.
Accordingly the collective excitations about this state are gapless
(as the Nambu-Goldstone modes),  
and the spectrum~(\ref{LWspec}) is anisotropic in ${\bf p}$.


\begin{figure}[tbp]
\begin{center}
\includegraphics[scale=0.23]{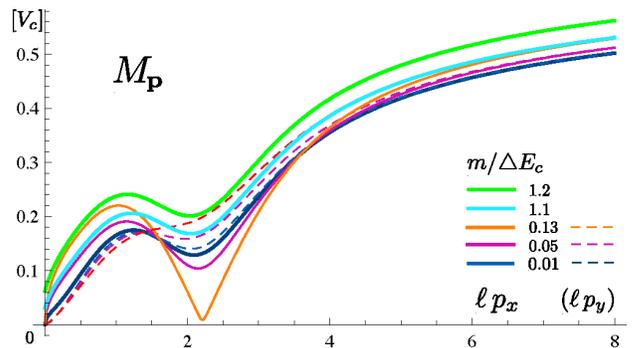}
\end{center}
\caption{
Excitation spectra $M_{\bf p}$, plotted in units of $V_{c}= \alpha/\epsilon_{b}\ell$
for $m/\triangle E_{c}$ = 0.01, 0.05, 0.13, 1.1 and 1.2.
The real curves refer to the profiles in $p_{x}$ at $p_{y}=0$ (i.e., normal to 
the in-plane field $E_{y}$) and dashed curves to those in $p_{y}$ at $p_{x}=0$. 
}
\end{figure}


The excitation spectrum $M_{\bf p}$ in general exhibits a local minimum 
(roton minimum) around $|{\bf p}|\ell \sim 2$; see Fig.~1~(a).
In the $0 < m < \triangle E_{c}$ range of case (ii),
the roton minimum changes critically with $m$:
The minimum in $p_{x}$ comes down, as $m$ is increased 
from zero, and touches zero (gap) 
at $m/\triangle E_{c} \approx 0.131$ 
(with $\theta_{\rm min} \approx 42.6^{\circ}$). The spectrum loses sense 
(becoming pure imaginary) until $m$ reaches 
$m/\triangle E_{c}\approx 0.869$ 
(with $\theta_{\rm min} \approx 137.4^{\circ}$) 
where the roton minimum reappears;
it then rises and returns to the the $m=0$ spectrum 
at $m=\triangle E_{c}$.
This peculiar feature inspires one to find an interesting structure
of $E(\theta)$,
\begin{equation}
E(\theta; m) = E(\pi - \theta; \triangle E_{c} - m),
\label{dualityrel}
\end{equation}
valid for ${\cal E}\not=0$ as well.  This implies that
the half-filled state realized at  $\theta_{\rm min}$ 
for a given $m < \triangle E_{c}/2$ 
is paired with the state realized 
with angle $\pi -\theta_{\rm min}$ at a larger intrinsic gap  
$\propto m'=\triangle E_{c} - m$.
Both the energy $E(\theta_{\rm min})$ and 
excitation spectrum $M_{\bf p}$
completely agree for this pair of states,
as seen from Eqs.~(\ref{FpGpWp}) and~(\ref{LMp}).
In particular, the filled $n=0$ level realized at $m=0$ and 
the filled $n=1$ level realized at $m=\triangle E_{c}$ 
share the same energy and collective excitations. 
We thus find a kind of (small-$m$/large-$m$) duality
in the $0 \le m \le \triangle E_{c}$ range.

In this $m$ range the texture state, taken to be 
homogeneous in space, acquires spontaneous
in-plane electric polarization $\propto \sin \theta_{\rm min}$.
The anomalous behavior of the roton spectrum 
about $m/\triangle E_{c} \sim 0.5$ or $\theta \sim \pi/2$,
mentioned above,
reflects a potential instability of the texture state 
due to spontaneous polarization. 
Such electrically polarized  {\sl homogeneous} configurations, 
unless polarization is relatively weak, are unstable 
against local charge inhomogeneities 
and would decay into {\sl inhomogeneous} configurations.

A local charge excess would align electric dipoles outward or inward,
and let them drift in a magnetic field.
One may thus imagine a picture of charged electric dipoles 
drifting around local charge centers (distributed 
randomly or in some patterns on the real sample and substrate), 
clockwise or anticlockwise depending 
on the sign of the local excess charge.
We speculate that the half-filled state 
in the realistic $0 < m < \triangle E_{c}$ range may 
form many such domains for stabilization.

Let us next set $m\rightarrow 0$, i.e., consider
the case of zero band gap $U=0$ and 
study the effect of ${\bf E}_{\parallel}$.
The in-plane field ${\cal E}= e \ell E_{y}/\sqrt{2} >0$ tilts the pseudospin 
toward $\theta = -\pi/2$ and competes with $\triangle E_{c}$
which favors $\theta=0$.
As a result, $\theta_{\rm min}$ varies
from $0$ to $-\pi/2$ as $E_{y}$ is increased.
The charge carriers thereby acquire a nonzero electric dipole moment 
$\propto \sin \theta$
and the pseudospin waves always have 
a finite excitation gap for ${\cal E} \not=0$; see Fig.~2.
For weak field $4\, {\cal E}/\triangle E_{c}\equiv R \ll 1$, one finds  
$\theta_{\rm min} \approx  - R^{1/3}$,
$E_{\rm min} \approx  - (3/16)\,  R^{4/3}\, \triangle E_{c}$,
and
$\kappa_{\zeta}^{2} \approx 3\kappa_{\eta}^{2} 
\approx (3/4)\,  R^{2/3}\, \triangle E_{c}$
so that the excitation gap grows as 
\begin{eqnarray}
M_{{\bf p}=0}  \approx (\sqrt{3}/2)\,
(4\,{\cal E}/\triangle E_{c})^{2/3} \, \triangle E_{c}.
\label{gapEsmall}
\end{eqnarray}
For larger ${\cal E} \gg \triangle E_{c}$ the gap rises almost linearly with ${\cal E}$,
\begin{equation}
M_{{\bf p}=0} \approx_{}2\, {\cal E} +  \triangle E_{c}/2,
\label{gapElarge}
\end{equation}
along with $\kappa^{2}_{\eta} \approx {\cal E}$ and 
$\kappa^{2}_{\zeta} \approx {\cal E} + \triangle E_{c}/2$.
As seen from Fig.~2, Eqs.~(\ref{gapEsmall}) and~(\ref{gapElarge})
combine to give a practically good description of 
the excitation gap over the entire range of ${\cal E}$; 
crossover takes place around ${\cal E}/\triangle E_{c} \sim 0.3$.
Note that the Coulomb correction significantly enhances 
the excitation gap; 
in particular, the gap rises 
prominently as ${\cal E}^{2/3}$ for a weak field.


\begin{figure}[tbp]
\begin{center}
\includegraphics[scale=.33]{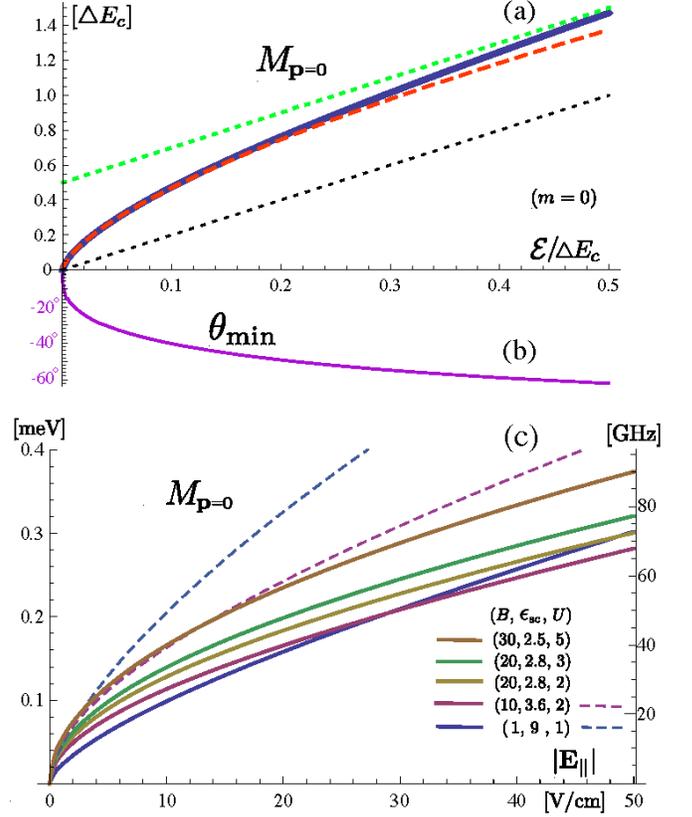}
\end{center}
\vskip-.7cm
\caption{
(a)~Excitation gap at zero momentum, plotted in units of
$\triangle E_{c}$ as a function of the in-plane field, 
${\cal E}/\triangle E_{c}$.
The red dashed curve refers to Eq.~(\ref{gapEsmall}) and
the green dotted line to the asymptotic form of 
Eq.~(\ref{gapElarge}).
(b)~Angle of inclination $\theta_{\rm min}$ in degrees.
(c)~Excitation gap $M_{{\bf p}=0}$, in units of meV and GHz, plotted 
as a function of the in-plane field $|E_{\parallel}|$ 
for some typical values of $(B[{\rm T}], \epsilon_{\rm sc}, U[{\rm meV}])$.
Dashed curves refer to the cases 
where the screening effect is turned off, 
$\epsilon_{\rm sc}\rightarrow 1$. }
\end{figure}

\begin{figure}[htbp]
\begin{center}
\includegraphics[scale=.43]{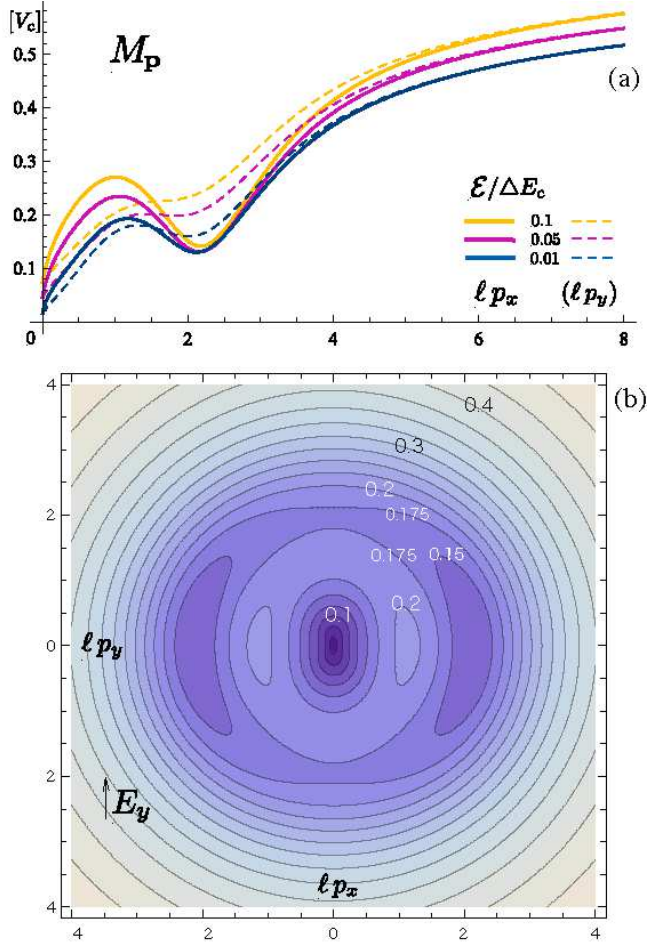}
\end{center}
\caption{(a) Excitation spectra $M_{\bf p}$, in units of 
$V_{c}= \alpha/\epsilon_{b}\ell$,
for ${\cal E}/\triangle E_{c} = 0.01$, 0.05 and 0.1.
The real curves refer to the profiles in $p_{x}$ and dashed curves
to those in $p_{y}$.  
(b) Contour plot of $M_{\bf p}$ for ${\cal E}/\triangle E_{c} = 0.02$.
}
\end{figure}


For ${\cal E}\not =0$ the excitation spectrum $M_{\bf p}$ 
is necessarily anisotropic, especially at low momenta,
and, as seen from a contour plot of $M_{\bf p}$ in Fig.~3, 
anisotropy of the roton minima already develops 
around ${\cal E}/\triangle E_{c} \sim 0.02$.
The spectrum recovers isotropy for $|{\bf p}|\ell \sim 5$ or larger and
the asymptotic spectrum is lifted roughly 
by the amount of the excitation gap,
as seen from the spectrum profiles 
in Fig.~3(a) and from Eq.~(\ref{Minf}).
Note that the roton minimum, unlike the $m\not = 0$ case, 
shows no sign of instability.

When both $m$ and ${\cal E}$ are present, their effects generally 
tend to add up.  The texture excitations always have a gap
and 
their potential instability, in a certain range 
about $m/\triangle E_{c} \sim 0.5$, weakens 
and eventually disappears with increasing ${\cal E}$;
the roton dip remains to be a local minimum as long as ${\cal E} >  m$.
We remark that the duality implied by Eq.~(\ref{dualityrel})
would hold in the presence of ${\cal E}$ as well.

In Fig.~2(c) we plot the excitation gap as a function of 
$|{\bf E}_{\parallel}|$ for some typical values of 
$B[{\rm T}]$ and $m\propto U[{\rm meV}]$.
With the effect of screening $\epsilon_{\rm sc}$ taken into account,
$M_{{\bf p}=0}$ falls in roughly the same frequency range of microwaves.

If ${\bf E}_{\parallel}$ is sufficiently strong, 
a sizable excitation gap $M_{{\bf p}=0}$
would arise, leading to an incompressible $\nu=2$ state.  
One could thereby observe the (spin-degenerate) $\nu=2$ Hall plateau 
with a suitably strong 
injected current for $U\not=0$ and $U=0$ as well.

Of the pseudospin $\propto {\bf n}$ the angle $\theta$ is 
related to the ratio in amplitude of the $n=0_{+}$ and $n=1$ modes 
while the angle $\phi$ is related to the relative phase between them.
One would now have control of mixing of the zero-mode levels 
by adiabatically changing the strength and direction 
of an injected current.

\section{current and Response}

In this section we study the response of the pseudospin waves 
in the $(0_{+},1)$ sector under uniform fields $B$ and $E_{y}$
to   a weak time-varying external field.
To this end we consider a weak vector potential 
${\bf A}(t)=(A_{x}, A_{y})$, which describes 
an external field $\sim \partial_{t}{\bf A}$ and, at the same time,
serves to probe the current.
For simplicity we take it to be spatially uniform.

The current operator $\delta H/\delta {\bf A}$ derives from two parts 
in $H$ of Eq.~(\ref{Hbilayer}).
One coming from ${\cal H}_{0}$ is the ordinary form of current, 
which has no projection to the $(0_{+},1)$ sector 
and induces transitions to other levels.
The other one, specific to bilayer graphene, derives 
from the $O(z U)$ part of ${\cal H}_{\rm as}$, 
and the relevant $O(A)$ portion of its $(0_{+},1)$ projection 
is written as 
\begin{eqnarray}
\bar{H}_{A} 
&=& -2\sqrt{2}\, m\,\rho_{0} e \ell\, 
( A_{x}\,S^{1}_{\bf p=0} - A_{y}\, S^{2}_{\bf p=0} ),
\end{eqnarray}
where $m= z\, U/4$.
Evaluating $\langle \tilde{G}| \bar{H}_{\bf A} | \tilde{G}\rangle$
yields an addition to ${\cal H}_{\rm coll}$
of the form   
\begin{eqnarray}
{\cal H}_{A} 
 &=& - \sqrt{2}\, e\ell\,
\{ c_{0}\, A_{x} + c_{x}\, A_{x}\zeta
+ c_{y}\, A_{y}\, \eta \}, \nonumber\\ 
c_{0}&=& m \sin\theta, c_{x}= m \cos \theta, c_{y}= m.
\label{HsubA}
\end{eqnarray}

One might read from ${\cal H}_{A}$ the current density 
carried by the zero-mode sector as 
$j_{x}^{\rm zm} =-\sqrt{2}\, \rho_{0}\, e\ell\, m \,\sin \theta_{\rm min}$
[which, for $V_{c}=0$, equals $- (\nu e^{2}/h)\, E_{y}$].
This, unfortunately, is not a complete amount of current yet. 
Inter-Landau-level transitions caused by the ordinary current 
induce some extra charge in the $(0_{+},1)$ sector.
In other words, the presence of ${\bf A}$ causes level mixing, 
which modifies the current within the $(0_{+},1)$ sector.
Such a modification was calculated earlier~\cite{KSsma} 
for standard QH systems and, in the present case, it is given by
\begin{eqnarray}
\delta \bar{\rho}_{\bf -p}
&\approx& (S^{0}_{\bf - p} - S^{3}_{\bf - p})\ u_{11}({\bf p}),
\nonumber\\ 
u_{11}({\bf p})\! 
&\approx& 
\!\!  2\, i\, e \ell^{2}\, 
(1- \ell^{2}{\bf p}^{2}/4)(p_{x}\, A_{y} -p_{y}\, A_{x}),
\end{eqnarray}
to  $O({\bf A})$, apart from terms of $O(\partial_{t}A_{i}/\omega_{c})$.
This $u_{11}({\bf p})$ represents charge accumulated 
in the $n=1$ level via the $n=1 \rightarrow \pm 2\rightarrow 1$ 
inter-Landau-level transitions.

The induced charge also carries current
within the $(0_{+},1)$ sector via the interaction 
\begin{equation}
\delta \bar{H}_{A} = {1\over{2}} \sum_{\bf p}v_{\bf p} 
:\{ \bar{\rho}_{\bf p}, \delta \bar{\rho}_{\bf -p}\}:
- \sum_{\bf p} e(A_{0})_{\bf p}\,  \delta \bar{\rho}_{\bf -p}.
\end{equation}
The current response is now calculated from
$\langle \tilde{G}| \bar{H}_{\bf A} + \delta \bar{H}_{A} |\tilde{G}\rangle$.
The result 
again takes the form of Eq.~(\ref{HsubA}),
with coefficients $(c_{0}, c_{x}, c_{y})$ modified as follows
\begin{eqnarray}
c_{0} &=& m\sin\theta + (1-\cos \theta)\, {\cal E} 
+ (V_{1}/32)\, \sin 2\theta,
\nonumber\\ 
&=& {\cal E}  - E'(\theta),
\label{czero} \\
c_{x} &=& m\cos\theta + \cdots = - E''(\theta),
\label{cx} \\
c_{y} &=& m + (V_{1}/16) (\cos\theta -1), \nonumber\\ 
&=& {\cal E}\cos \theta/\sin\theta  - E'(\theta) /\sin\theta.
\label{cy} 
\end{eqnarray} 
With $E'(\theta) \rightarrow 0$, Eq.~(\ref{czero}) verifies
that the pseudo-zero-mode sector carries 
the {\sl correct} amount of Hall current
(in response to a uniform field ${\cal E}$) 
with conductance $\sigma_{xy} = - \nu\, e^{2}/h$.

The Hamiltonian $\bar{H}_{A} +\delta \bar{H}_{A}$ governs 
the microwave response 
of the pseudo-zero-mode sector. 
It is combined with ${\cal H}_{\rm coll}$ to yield the source term 
$\Phi\, \{ 2 c_{y}\,(\Gamma_{\zeta})^{1/2} A_{y} 
-  c_{x}\, (\Gamma_{\zeta})^{-1/2}\, \partial_{t}A_{x}$
for $L_{\Phi}$ in Eq.~(\ref{LMp}).
Solving for the stationary action then yields a response
of the form  
$\sim A_{y} (\cdots) \partial_{t}A_{x}$.
From this one can read off the optical Hall conductance due to virtual transitions 
within the $(0_{+}, 1)$ sector,  
\begin{equation}
\triangle\sigma_{xy}(\omega) = - {\nu\, e^{2}\over{2\pi \hbar}}
\cos \theta_{\rm min}\, 
{M_{\bf 0}^{2}\over{M_{\bf 0}^{2} - \omega^{2}}},
\end{equation}
which is significantly peaked around 
$\omega \sim M_{\bf 0}\equiv M_{{\bf p}=0}$,
the pseudospin-wave gap.

The collective excitations within the $(0,1)$ sector thus contribute 
the $\cos\theta$ portion [$\triangle\sigma_{xy}(\omega\rightarrow 0)$]
of the Hall conductance $\sigma_{xy}$ while the remaining $(1-\cos \theta)$
portion of $\sigma_{xy}$ essentially comes from
the $n=1 \rightarrow \pm 2\rightarrow 1$ virtual transitions 
with larger gaps $\sim \sqrt{2}\, \omega_{c}
\gg M_{\bf 0}$.
These two components are distinguishable via microwave or light response.

With disorder taken into account, the diagonal conductivity 
$\triangle\sigma_{xx}(\omega)$ also is significantly peaked 
around $\omega \sim M_{\bf 0}$, and 
$M_{\bf 0}$  varies critically with ${\cal E}$
or by an injected current, as we have seen in Fig.~2.
Microwave or infrared experiments,~\cite{bilayerExp}
via absorption, reflection or conductance fluctuation,
would provide a direct means to explore 
such unique dynamics of the pseudo-zero-modes.

\section{summary and Discussion}

Zero-mode Landau levels, specific to graphene in a magnetic field, 
are very special.  Their presence has a topological origin 
in the chiral anomaly.
They show quite unusual dielectric response that reflects 
quantum fluctuations of the vacuum state (the Dirac sea).
Bilayer graphene supports eight such zero-mode levels 
which, unlike in monolayer graphene, involve two different 
orbital indices $n=0, 1$.
As a tunable band gap develops, four of them at one valley 
are isolated from the others at another valley 
and remain nearly degenerate although its fine structure
sensitively depends, via mixing of zero-modes, on the environment.

In this paper we have studied the effects of 
an external field and the Coulomb interaction 
on such an isolated zero-mode quartet.
This pseudo-zero-mode sector,
especially at half filling, supports, via orbital mixing,
quasiparticles with charge and electric dipole,
which give rise to characteristic collective excitations, 
pseudospin waves.
We have constructed a low-energy effective theory 
of pseudospin waves with general pseudospin textures and 
noted a duality [Eq.~(\ref{dualityrel})]
in the excitation spectrum.
The excitation gap at zero momentum turns out to be generally small, 
reflecting the intrinsic degeneracy of the pseudo-zero-mode sector,
and the Coulomb exchange energy works 
to enhance the effect of the in-plane field on the gap.
This means that the gap is tunable by an in-plane field or 
by an injected current; the mixing of the zero-modes
(i.e., relative phase and magnitude) is also externally controllable 
to some extent.

The pseudo-zero-mode sector of bilayer graphene 
is particularly suited for exploring coherence phenomena.
This is because it essentially resides on the same layer
so that, unlike in ordinary bilayer QH systems, 
there is no cost of interlayer capacitance energy for it.

An experimental signature of the field-induced gap is 
to observe the quantum Hall effect 
with an injected current; one would be able 
to resolve the $\nu= \pm 2$ Hall plateaus 
(or the spin-resolved $\nu=\pm1$ plateaus)
using a suitably strong current.

The collective excitations within the pseudo-zero-mode sector 
also carry a considerable portion of the total current.
A direct study of the excitation gap $M_{{\bf p}=0}$ and 
its field dependence by microwave absorption or reflection 
would clarify the unique controllable features of 
the pseudo-zero-mode sector in bilayer graphene, 
and, in addition, the effect of screening 
on the Coulomb correlation energy $\triangle E_{c}$ 
due to vacuum polarization.

\acknowledgments

The author wishes to thank T. Morinari for useful discussions. 
This work was supported in part by a Grant-in-Aid for Scientific Research
from the Ministry of Education, Science, Sports and Culture of Japan 
(Grant No. 17540253).

\appendix

\section{Static structure factors}

In this appendix we outline the derivation of the pseudospin static structure 
factors $\langle S^{a}_{\bf p}S^{b}_{\bf q}\rangle$ [in Eq.~(\ref{SpSq})]
for the half-filled pseudo-zero-mode levels $|G\rangle$
with $\langle S^{a}_{\bf 0}\rangle = (N_{e}/2)\, n^{a}$ 
pointing in a general direction ${\bf n}=\{n^{a}\}$ in pseudospin space. 
Let us first note that, 
when only the $n=0_{+}$ level is filled, i.e., for $n^{3}=1$ polarization,
these structure factors are readily calculated:
Filling the $n=0_{+}$ level and leaving the $n=1$ level empty 
immediately imply that  
$\langle\, S^{\mu}_{\bf p}S^{\nu}_{\bf q} \rangle
= (\rho_{0}/2)^{2}\, 
\delta_{\bf p,0}\, \delta_{\bf q,0}$ for  $\mu,\nu \in (0,3)$,
and $\langle \,:\! S^{a}_{\bf p} S^{b}_{\bf q}\!: \,\rangle = 0$
for $a,b \in (1,2)$. 
One may then note the algebraic relation 
\begin{equation}
S^{a}_{\bf p} S^{b}_{\bf q}
= e^{f_{\bf p q}} {1\over{2}}\, ( \delta^{ab}\, S^{0}_{\bf p+q} 
+ i\, \epsilon^{abc}\, S^{c}_{\bf p+q} ) + : S^{a}_{\bf p} S^{b}_{\bf q} : 
\end{equation}
with $f_{\bf p q} = 
\ell^{2}({\bf p}\cdot {\bf q} - i\, {\bf p} \times {\bf q})/2$,
and determine, e.g.,  
$\langle\, : \! S^{3}_{\bf p}S^{3}_{\bf -p}\!\! :\rangle 
 = (N_{e}/4)  (\rho_{0}\delta_{\bf p,0}- \gamma_{\bf p}^{2})$ and
$\langle S^{1}_{\bf p} S^{1}_{\bf -p}\rangle 
= (N_{e}/4)\, \gamma_{\bf p}^{2}$, 
with $\gamma_{\bf p} = e^{- \ell^{2}{\bf p}^{2}/4}$.

The structure factors for the half-filled state 
with a general pseudospin polarization
${\bf n}$ are obtained from these $n^{3}=1$ structure factors 
by a suitable rotation in pseudospin space.
Note first that the $n^{3}=\sigma^{3}=\pm 1$ eigenspinors $|\pm 1 \rangle$ 
of the Pauli matrix $\sigma^{3}$ are rotated by angle 
$(\theta, \phi)$ to form the $\sigma^{a} = \pm n^{a}$ eigenspinors 
$U|\pm 1 \rangle$ with $U =e^{- i \phi \sigma^{3}\!/2}\,
e^{- i \theta \sigma^{2}\!/2}$.
Accordingly we decompose the zero-mode field 
$\Psi=(\psi_{0_{+}}, \psi_{1} )^{\rm t}$
[defined in Eq.~(\ref{Smup})] into 
the $\sigma^{a} = \pm n^{a}$ eigenmodes $\Psi'$ 
by writing $\Psi = U \Psi'$.

On substitution $\Psi = U \Psi'$, $S^{a}_{\bf p}$ are rewritten
as linear combinations of the pseudospin operators 
$S'^{a}_{\bf p} \sim \Psi'^{\dag}  (\sigma^{a}/2)\, e^{i {\bf p \cdot r}}\Psi'$ 
composed of $\Psi'$; $S^{0}_{\bf p}=S'^{0}_{\bf p}$.
One can then calculate $\langle S^{\mu}_{\bf p}S^{\nu}_{\bf q}\rangle$ 
for general $n^{a}$ from 
the structure factors $\langle S'^{\mu}_{\bf p}S'^{\nu}_{\bf q}\rangle$
for the $n^{3}=1$ state.
The result is summarized in Eq.~(\ref{SpSq}).

Note, in particular, that 
$S^{a}_{\bf p} = n^{a} S'^{3}_{\bf p}+ \cdots$.  
This yields $\langle S^{a}_{\bf p}\rangle = n^{a}\langle S'^{3}_{\bf p}\rangle 
= n^{a} ( \rho_{0}/2)\, \delta_{\bf p,0}$ and tells us  that 
the normal-ordered factors take particularly simple form
$\langle\,:~\! S^{a}_{\bf p}S^{b}_{\bf q}\!: \rangle
= n^{a} n^{b} \langle\, :\! S'^{3}_{\bf p}S'^{3}_{\bf q}\! : \rangle$,
or
\begin{equation}
\langle\, :\! S^{\mu}_{\bf p}S^{\nu}_{\bf -p}\!\!: \rangle
= -n^{\mu}n^{\nu} (N_{e}/4) ( \gamma_{\bf p}^{2} - \rho_{0}\delta_{\bf p,0}),
\end{equation}
with $n^{0}=1$, as quoted in Eq.~(\ref{normSaSb}).

\section{collective excitations}

In this appendix we outline the derivation of ${\cal H}_{\rm coll}$
in Eq.~(\ref{Hcoll}). 
Let us first consider the contribution from the Coulomb interaction,
${1\over{2}}\, \sum_{\bf p}v_{\bf p} J_{\bf p}$
with $J_{\bf p}=\langle G | (\bar{\rho}_{\bf p})^{\cal O}\,
(\bar{\rho}_{\bf -p})^{\cal O} |G \rangle$ and
$(\bar{\rho}_{\bf p})^{\cal O} = e^{i{\cal O}}\bar{\rho}_{\bf p} 
e^{-i{\cal O}}$.
Expanding  $(\bar{\rho}_{\bf p})^{\cal O}$ 
in powers of ${\cal O} \propto \Omega$ 
by repeated use of the $W_{\infty}$ algebra~(\ref{chargealg}) and 
subsequently substituting the structure factors in Eq.~(\ref{SpSq})
allow one to evaluate $J_{\bf p}$.

The $O(\Omega)$ term is thereby written as
\begin{eqnarray}
J_{\bf p}^{(1)} 
&=& - \rho_{0}\, \gamma_{\bf p}^{2}\, \Omega^{a}_{\bf k=0}\,
\epsilon^{abc}
w^{b}_{\bf p}\,w^{\beta}_{\bf -p}\, s^{\{c,\beta\}},
\nonumber\\ 
&=&  \rho_{0}\, \gamma_{\bf p}^{2}\, \Omega^{2}_{\bf 0}\,
( |w^{1}_{\bf p}|^{2} -|w^{3}_{\bf p}|^{2})\, s^{\{1,3\}} ,
\end{eqnarray}
under a symmetric integration over ${\bf p}$; 
$s^{\{c,\beta\}} \equiv s^{c \beta} + s^{\beta c}$ for short.
(Here, we have employed the convention $\phi=0$ 
and $n^{2}=0$, as remarked in the text.)

Similarly, the $O(\Omega \Omega)$ term is written as
\begin{eqnarray}
J_{\bf p}^{(2)} 
&=& \rho_{0}^{2}\, \gamma_{\bf p}^{2}\, 
  |\epsilon^{abc}\,  n^{a}w^{b}_{\bf p}\, \Omega^{c}_{\bf -p}|^{2}
\nonumber\\ 
&&+  {\rho_{0}\over{2}}\,\gamma_{\bf p}^{2} \sum_{\bf k} \cos^{2}
 \Big({1\over{2}}\, \ell^{2} {\bf k}\!\times\! {\bf p}\Big)\, (N_{1} + N_{2})\nonumber\\ 
&& +  {\rho_{0}\over{2}}\,\gamma_{\bf p}^{2} \sum_{\bf k}  \sin^{2}
 \Big({1\over{2}}\, \ell^{2} {\bf k}\!\times\! {\bf p}\Big)\, (N_{3} + N_{4})\nonumber\\ 
&& -{\rho_{0}\over{4}}\, \gamma_{\bf p}^{2}\,  
\sum_{\bf k} \sin ( \ell^{2} {\bf k}\!\times\! {\bf p})\, (2 N_{5} + N_{6}),
\end{eqnarray}
with 
\begin{eqnarray}
N_{1}&=& \Omega^{\alpha}_{\bf -k}\, \Omega^{a}_{\bf k}\, 
w^{\beta}_{\bf -p}\, w^{b}_{\bf p}\, 
\epsilon^{\alpha\beta\gamma}\, \epsilon^{abc} s^{\{\gamma, c\}},
\nonumber\\ 
N_{2}&=& - \Omega^{\alpha}_{\bf -k}\, \Omega^{a}_{\bf k}\, 
w^{b}_{\bf p}\, w^{\beta}_{\bf -p}\, \epsilon^{abj} \epsilon^{\alpha c j}\, 
s^{\{c,\beta\}}, \nonumber\\ 
N_{3} &=& \Omega^{\alpha}_{\bf -k}\, \Omega^{a}_{\bf k}\, 
|w^{0}_{\bf p}|^{2}\, s^{\{\alpha, a\}},\nonumber\\ 
N_{4} &=& - \Omega^{\alpha}_{\bf -k}\, \Omega^{a}_{\bf k}\, w^{a}_{\bf p}\,w^{\beta}_{\bf -p}\, 
   s^{\{\alpha,\beta\}}, \nonumber\\ 
N_{5} &=& \Omega^{\alpha}_{\bf -k}\, \Omega^{a}_{\bf k}\, w^{0}_{\bf p} \, 
 w^{\beta}_{\bf -p}\, \epsilon^{\alpha\beta\gamma}
  s^{\{\gamma, a\}}, \nonumber\\ 
N_{6} &=& \Omega^{a}_{\bf -k}\, \Omega^{b}_{\bf k}\, \epsilon^{abc}\, w^{0}_{\bf p}\, 
w^{\beta}_{\bf -p}\,  s^{\{c,\beta\}},
\end{eqnarray}
where ${\bf k}\!\times\! {\bf p}= k_{x}p_{y}- k_{y}p_{x}$.

One can evaluate $\sum_{\bf p}v_{\bf p} J_{\bf p}$ 
by integrating over ${\bf p}$ and leaving the ${\bf k}$
integration as it is. 
One may express the sines and cosines
in terms of $e^{\pm i\ell^{2} {\bf p}\times {\bf k}}$.
Integration over ${\bf p}$ is then carried out 
as a Fourier transform of the form 
$\sum_{\bf p}v_{\bf p} e^{- \ell^{2}{\bf p}^{2}/2 
+ i{\bf p \cdot x}}\, ({\rm powers\ of \ } p_{i})$
with ${\bf x} \rightarrow  \ell^{2}(\! \times {\bf k}) 
\equiv \ell^{2}\, (k_{y}, -k_{x})$.

The rest of terms in ${\cal H}_{\rm coll}$ are obtained 
via the induced pseudospin to $O(\Omega \Omega)$ 
\begin{eqnarray}
\langle (S^{a}_{\bf p})^{\cal O} \rangle 
&=& {\rho_{0}\over{2}}\, \gamma_{\bf p} \Big[ n^{a}\, \delta_{\bf p, 0} 
+    \epsilon^{abc}\, \Omega^{b}_{\bf p}\, n^{c}
 \nonumber\\ 
&&
- {1\over{2}}\,\{
n^{a}\, (\Omega^{b}, \Omega^{b})_{\bf p}
-(\Omega^{a}, \Omega^{b})_{\bf p}\, n^{b} \}\Big],\ \ \
\label{inducedpspin}
\end{eqnarray}
where
$(\Omega^{a}, \Omega^{b})_{\bf  p} \equiv \sum_{\bf k}
\cos (\ell^{2} {\bf k}\!\!\times\!\! {\bf p}/2)\, 
\Omega^{a}_{\bf -k+p}\, \Omega^{b}_{\bf k}$ for short.
This, in particular, is used to evaluate the contribution 
$\langle \triangle^{\cal O} \rangle$ from $\triangle$  
in Eq.~(\ref{triangle}).
Somewhat tedious calculations along these lines 
eventually lead to ${\cal H}_{\rm coll}$
in Eq.~(\ref{Hcoll}).

\section{integrals}

The integrals appearing in Eq.~(\ref{integrals}), 
apart from their overall factors,
are expressed in terms of the modified Bessel functions
\begin{equation}
\int\! dz\,  e^{-z^2/2}\,(\cdots) 
= e^{- q^{2}/4} \sqrt{\pi\over{2}}\, 
\Big[ c_{0}I_{0}(q^{2}/4) + c_{1}I_{1}(q^{2}/4) \Big],
\end{equation}
with coefficients
\begin{eqnarray}
(c_{0}, c_{1}) &=& (1/4)(2+q^{2}, -q^2)\  {\rm for\ } \xi_{q},
\nonumber\\ 
(c_{0}, c_{1}) &=& (1/2)(q^{2}, -2 - q^2)\  {\rm for\ }  \lambda_{q}, 
\nonumber\\ 
(c_{0}, c_{1}) &=& (q^2/16)(-2+q^{2}, - q^2)\  {\rm for\ } b_{q},
\nonumber\\ 
(c_{0}, c_{1}) &=& -(q/8)(1+q^{2}, -3- q^2)\  {\rm for\ }\tau_{q}.
\end{eqnarray}



\begin{thebibliography}{99}
\bibitem{fnOmega} 
To recover the $\phi$ degree of freedom one may simply replace
$(\Omega^{1}, \Omega^{2})$  with
$\Omega'^{1} = \Omega^{1}\, \cos \phi +\hat{\Omega}^{2}\, \sin \phi$ 
and 
$\Omega'^{2} = -\Omega^{1}\, \sin \phi +\Omega^{2}\, \cos \phi$, 
and rotate $(p_{y}, p_{x})$ analogously.
\bibitem{fnvac} 
With only the effect of vacuum polarization, 
calculated in Ref.~[\onlinecite{MS}], is taken into account,
one finds $\epsilon_{\rm sc}\approx$ 9, 3.6, 2.8, and 2.5 
for $B[{\rm T}]$=1, 10, 20, and 30,
respectively.
\end{thebibliography}
\end{document}